\documentclass[fleqn,10pt]{wlscirep}

\title{Two-dimensional electronic spectra of the photosynthetic apparatus of green sulfur bacteria}

\author[1,2,*,+]{Tobias Kramer}
\author[1,+]{Mirta Rodr{\'i}guez}
\affil[1]{Konrad-Zuse-Zentrum f\"ur Informationstechnik Berlin, 14195 Berlin, Germany} 
\affil[2]{Department of Physics, Harvard University, 17 Oxford St, Cambridge, 02138 Massachusetts, U.S.A.} 
\affil[+]{these authors contributed equally to this work}
\affil[*]{kramer@zib.de}

\begin{abstract}
Advances in time resolved spectroscopy have provided new insight into the energy transmission in natural photosynthetic complexes.
Novel theoretical tools and models are being developed in order to explain the experimental results.
We provide a model calculation for the two-dimensional electronic spectra of {\it Cholorobaculum tepidum} which correctly describes the main features and transfer time scales found in recent experiments.
From our calculation one can infer the coupling of the antenna chlorosome with the environment and the coupling between the chlorosome and the Fenna-Matthews-Olson complex.
We show that environment assisted transport between the subunits is the required mechanism to reproduce the experimental two-dimensional electronic spectra.
\end{abstract}

\begin{document}

\flushbottom
\maketitle
\thispagestyle{empty}

\section*{Introduction}

\begin{figure}[t]
\begin{center}
\includegraphics[height=0.5\textwidth]{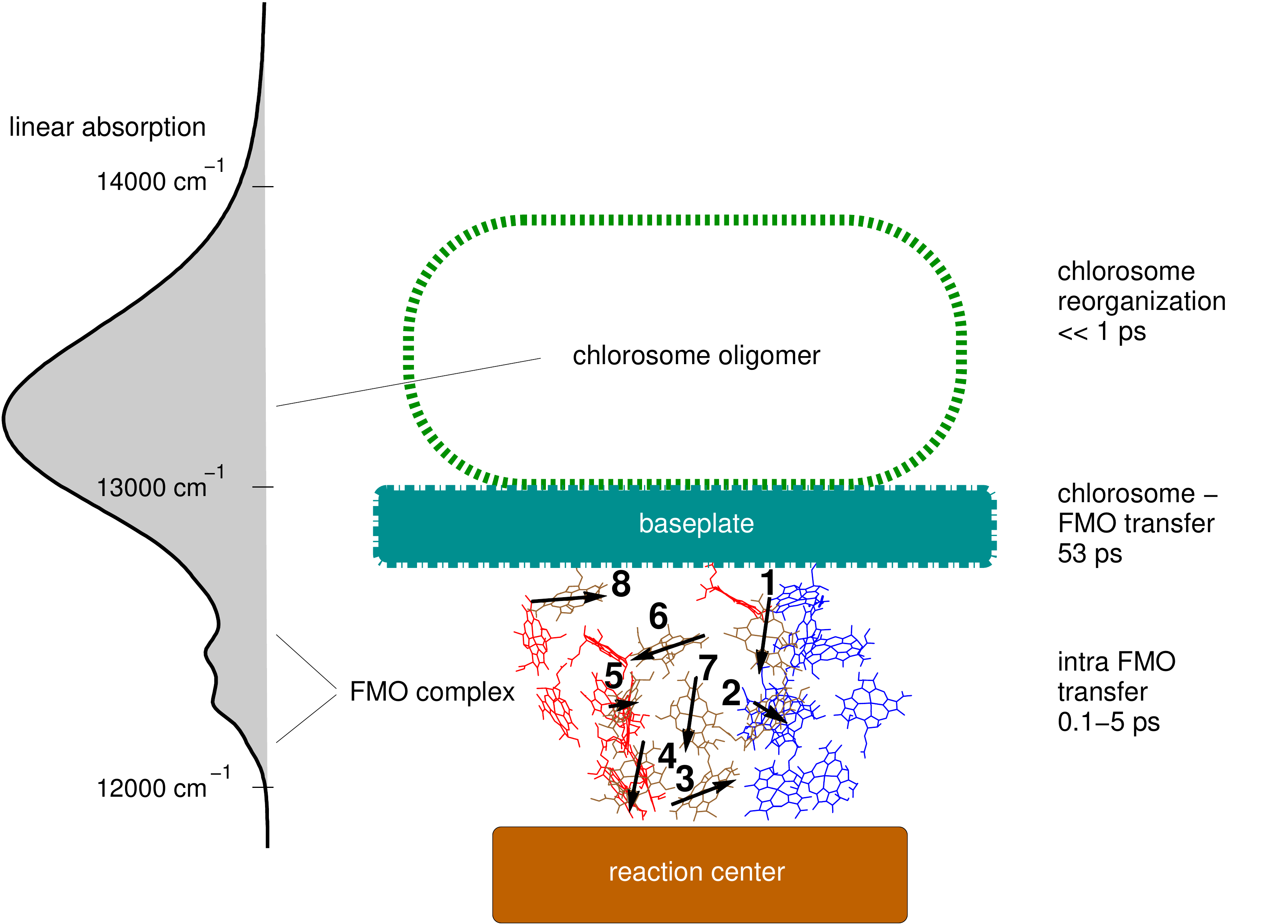}
\end{center}
\caption{
Linear absorption spectra and schematics of the photosynthetic apparatus of {\it C.~tepidum}.
The antenna chlorosome, baseplate, FMO complex, and reaction center are arranged as an energetic funnel, as revealed in the linear absorption spectrum.
Dipole directions of eight BChls~a within a FMO monomer are indicated by black arrows.
The linear absorption spectra has been calculated at temperature $300$~K with parameters from Table~\ref{tab:param} (see Methods).
\label{fig:ctepschema}}
\end{figure}

Recent experiments, for the first time, track the energy flow through the whole photosynthetic apparatus of {\it Cholorobaculum tepidum} ({\it C.~tepidum}) \cite{Dostal2016}.
{\it C.~tepidum} are green sulfur bacteria that perform anoxygenic photosynthesis under the lowest light intensities known \cite{Blankenship2014}.  
Their photosynthetic apparatus consists of specialized antenna structures known as chlorosome, which are linked via the Fenna-Matthews-Olson (FMO) complexes to the reaction centers  (RC) where oxidation takes place.
The chlorosome is a large ellipsoidal structure formed by self-assembled aggregates of $\sim 10^5$ bacteriochlorophyll (BChl) c pigments, organized into large oligomers with relatively little protein involvement \cite{Orf2013}.
The chlorosome rests upon a paracrystalline aggregate of BChl~a pigments and CsmA proteins, the baseplate. The homotrimeric BChl~a pigment-protein FMO complex sits between the chlorosome baseplate and the RC \cite{Wen2009} (see Fig.~\ref{fig:ctepschema}).
The FMO complex consists of three monomeric units, each composed of seven BChl~a pigments  \cite{Li1997} and an additional satellite BChl~a pigment later discovered \cite{Wen2011} surrounded by proteins.  On a larger scale, recent electron microscopy studies propose the existence of several supercomplexes formed by two RC and four FMO trimers attached to each chlorosome antenna \cite{Bina2016}.

Optical measurements and atomistic-based calculations together with mass spectroscopy data have determined an energy funnel structure resulting in directed energy flow in {\it C.~tepidum} from the antenna to the RC \cite{Wen2009}.
Light absorbed at high energy in the antenna system moves within the photosynthetic unit as electronic excitation of the individual pigments, delocalised into exciton states.
The coupling of the exciton to the molecular motion and vibration leads to thermal dissipation, which eventually directs the exciton to the RC located spatially next to the  energetically lowest exciton state \cite{Blankenship2014}.
Already within a single pigment, reorganization processes lower the energy of the excited state due to the shift of the nuclear coordinates to the equilibrium positions \cite{Ishizaki2009c}.
The transfer time from the antenna through the FMO complex to the reaction center depends critically on the strength of the inter-pigment dipole-dipole couplings, in addition to the vibrational dissipation. 
For achieving efficient transport, the transfer time must be short in comparison with the loss rate due to exciton decay \cite{Plenio2008,Rebentrost2009a,Kreisbeck2011}.

The composition of the light harvesting complexes is determined using X-ray, mass spectroscopy and electron microscopy experiments. 
The  structure of the FMO protein \cite{Fenna1975,Olson2004}, the chlorosome \cite{Orf2013} and more recently the RC \cite{Hauska2001} and the  baseplate \cite{Nielsen2016} have been characterised using these methods. 
Fitting of optical spectra and direct structure-based quantum-chemistry calculations have been used to obtain the coupling and relaxation rates within the FMO \cite{Adolphs2006a} and antenna \cite{Montano2003, Prokhorenko2003} subsystems. 
Recent optical experiments analyse the time-dependent tranfer between the chlorosome and the baseplate \cite{Dostal2014, Kell2015} and from the FMO to the RC \cite{He2015}. 
Due to the high computational complexity and missing parameters, theoretical attempts to describe the transfer within the whole photosynthetic system are limited \cite{Huh2014}.
Using scattering resistant two-dimensional electronic spectroscopy (2DES), Dostal {\it et al} show that across the entire apparatus of {\it C.~tepidum} the energy flows at a timescale of tens of picoseconds \cite{Dostal2016}.
This relatively slow transfer is augmented by a faster transfer within the subunits on a timescale of sub picoseconds to a few picoseconds. 
2DES  is a very powerful experimental technique with high temporal and spectral resolution that has been successfully applied to study the energy-transfer pathways in photosynthetic complexes, in particular for the FMO system \cite{Brixner2005, Engel2007a}.
In contrast to absorption or fluorescence measurements, 2DES  provides the full correlation map between the excitations and the probing wavelengths as a function of time after initial light absorption \cite{Mukamel1995,Cho2005,Nuernberger2015}.
This allows one to infer the couplings between the electronic transitions and additionally to probe the finite time scale of the transfer and reorganization processes (Fig.~\ref{fig:ctepschema}).

In this paper we present a theoretical modelling of the joint chlorosome and FMO system, that allows us to reproduce the experimental 2DES \cite{Dostal2016}.
We use the exact hierarchical equations of motion (HEOM) formalism \cite{Tanimura1989,Ishizaki2009,Kreisbeck2011} that considers the exciton and the vibrational environment on the same footing.
Moreover, HEOM is capable of treating the different reorganisation energies in the different subunits found in the photosynthetic apparatus.
In particular the inclusion of the tightly packed antenna chlorosome leads to a larger reorganisation energy \cite{Fujita2012} ($\sim 300$~cm$^{-1}$), which is almost tenfold increased compared to the FMO subunit \cite{Adolphs2006a}.
By varying the model parameters we derive the effective coupling and connectivity between the antenna chlorosome and the FMO protein which matches the experimental data.
In addition we extract the relative orientation of the antenna chlorosome with respect to the FMO trimer.

\section*{Results}

The energy funnel structure of the {\it C.~tepidum} is revealed in the calculated absorption spectra shown in Fig.~\ref{fig:ctepschema}. 
It presents a strong absorbing chlorosome peak at high energy followed by a sequence of lower peaks located in the FMO region.  
Both, the baseplate and the RC, are low absorbing subunits and are not distinguishable in the experimental absorption spectra and 2DES signal of the whole photosynthetic unit \cite{Dostal2016}. 
We have thus excluded them from our model calculations. 
From the line shapes and positions of the FMO complex and chlorosome peaks in the linear absorption we infer that the antenna chlorosome can be modeled as an effective exciton with strong environmental coupling (parameters in Table \ref{tab:param}). 

\begin{figure}[t]
\begin{center}
\includegraphics[width=0.9\textwidth]{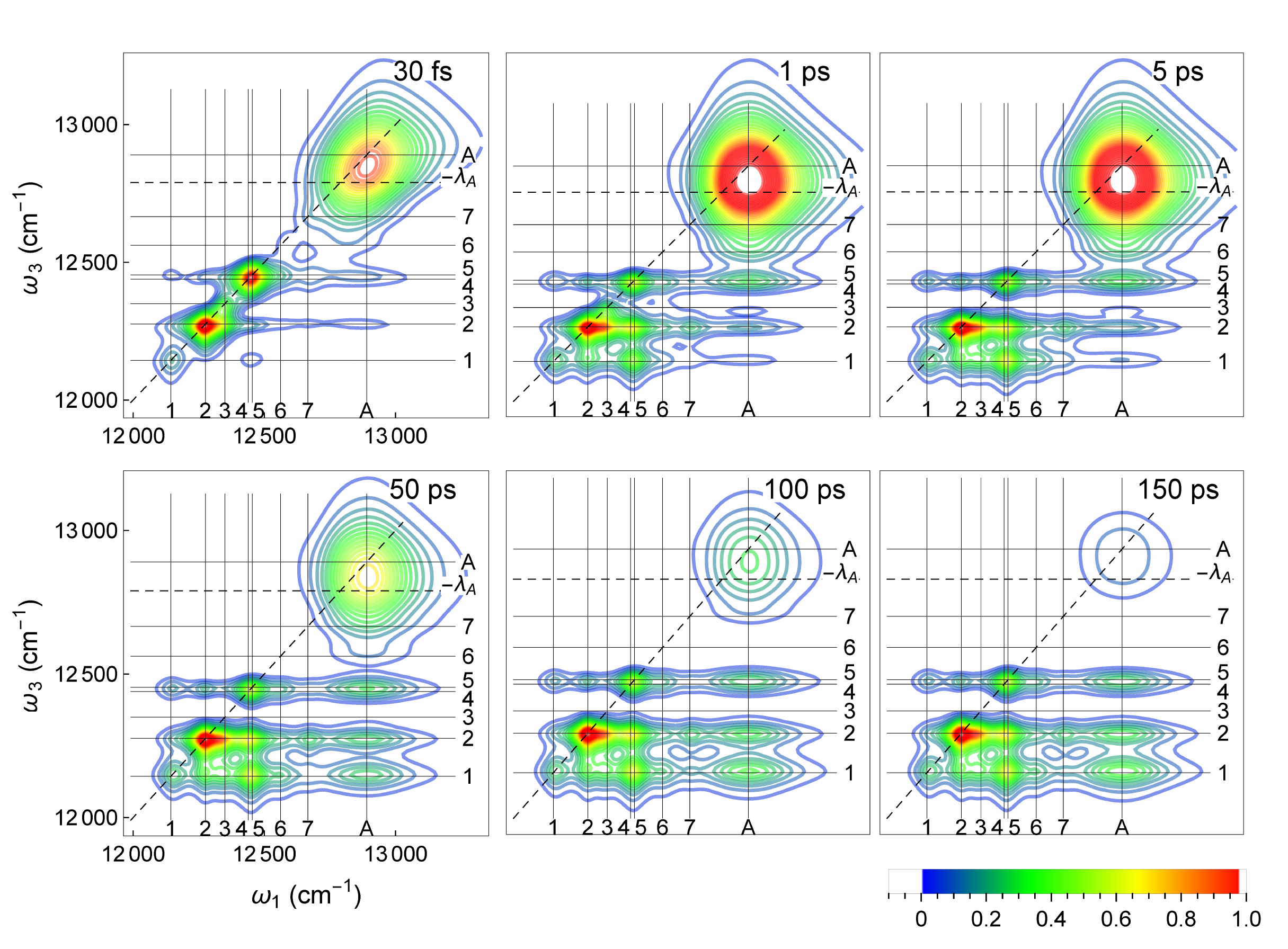}
\end{center}
\caption{Computed 2DES of {\it C.~tepidum} 
for delay times $\tau=1-150$~ps at temperature $150$~K including all rephasing and non-rephasing contributions with excitation frequency $\omega_1$ and emission frequency $\omega_3$.
We use the coupling and reorganisation rate parameters in Tab.~\ref{tab:param} (see Methods). 
\label{fig:all2DES}}
\end{figure}

The linear absorption signal yields the dipole strengths and approximate site energies, but does not provide insight into the exciton dynamics and reorganization shifts.
These processes are unfolded with the 2DES (Fig.~\ref{fig:all2DES}).
In 2DES a sequence of three laser pulses creates coherences between the ground state and between the exciton states, which are read out by a fourth pulse.
The first and last time intervals between the pulses are converted by a Fourier transform to the frequency domain, with $\omega_1$ and $\omega_3$ denoting the excitation and emission frequencies, respectively.
The remaining interval between second and third pulse sets the delay time $\tau$.
In 2DES the energy transfer is studied by comparing the peaked signals in the frequency domain at different delay times.
To avoid the overexposure of the FMO complex by the more absorptive chlorosome subunit, the experimental 2DES are measured with a different laser spectrum compared to the linear absorption spectra \cite{Dostal2016}.
We do account for the modified laser spectrum by shifting the  position of the chlorosome from $S_A^0$ to the peak position $S_A$ resulting from the convolution with the 2DES laser spectrum.
The weaker laser intensity in the chlorosome region is taken into account by an effectively reduced dipole strength, Tab.~\ref{tab:param}.

The panels in Fig.~\ref{fig:all2DES} show the results of our 2DES calculation at $T=150$~K for  delay times ranging from 30~fs to 150~ps.
The signals are peaked around $(\omega_1,\omega_3)  = (i,j)$ , where $i=1,\ldots,7,A$ denote the energies of the FMO exciton and antenna chlorosome  states.
For the shortest delay time $\tau=30$~fs the intensity is distributed along the diagonal line 
$\omega_1=\omega_3$.
The diagonal peaks indicate the energies where the light was absorbed.
With increasing delay time $1-150$~ps the chlorosome peak $(A,A)$ weakens in amplitude and off-diagonal cross-peaks $(i,j)$ appear below it.
The redistribution of the diagonal peaks into cross peaks is the indication of the excitation flow from the antenna unit to the FMO complex.
This process continues until the chlorosome peak fades around $\tau \sim 200$~ps \cite{Dostal2016}.
At $1$~ps, the diagonal peaks in the FMO transitions $(i,i)$, $i=2$-$5$, decay into off-diagonal peaks, indicating energy transfer to the energetically lower FMO states.
The antenna peak becomes broader and shifts to a lower $\omega_3$ frequency due to the chlorosome reorganization processes.
At $5$~ps the intra-FMO complex flows have reached a steady state, while the 
chlorosome peak $(A,A)$ still decays and transfers energy into the FMO.
Cross-peaks $(A,1)$ and $(A,2)$ start to appear and intensify at 50~ps, 
while cross-peaks intensities $(A,4)$ and $(A,5)$ remain almost constant.
Starting from $100$~ps, when the delay time $\tau$ exceeds the chlorosome thermalization time, the cross peak at the intersection of  the antenna and the energetically lowest FMO state $1$ (situated close to the RC on FMO BChl~3 and 4) dominates the $(A,i)$ signals. 
In this regime, we only observe variations in the 2DES due to the slow decay into the thermal equilibrium state.

To ensure numerical convergence of the HEOM method, which increases with increasing temperature \cite{Ishizaki2009h}, we have computed the 2DES at temperature $150$~K, while the experimental 2DES are recorded at $77$~K.
Compared to the experimental values, we obtain a lower $(A,1)$ peak intensity due to the different thermal populations of the lowest eigenstate of the FMO complex ($0.6$  vs.\ $0.9$).
This is in agreement with the temperature dependent experimental results in Fig.~S1 of Ref.~\citenum{Thyrhaug2016}.
The time scales inferred from the 2DES are only minimally affected by the change in temperature, while the reorganization energy has a more dramatic effect (see below).
In our calculations we have used a small dipole-dipole coupling $J_{A}=3$~cm$^{-1}$  between the antenna and the BChl~$1$ and $6$.
This coupling rate, together with the reorganisation energy and the onsite energy landscape are responsible for the transfer time of the energy flow between the chlorosome antenna and the FMO complex subunits.

\subsection*{Model parameters}

\begin{figure}[t]
\begin{center}
\includegraphics[width=0.99\textwidth]{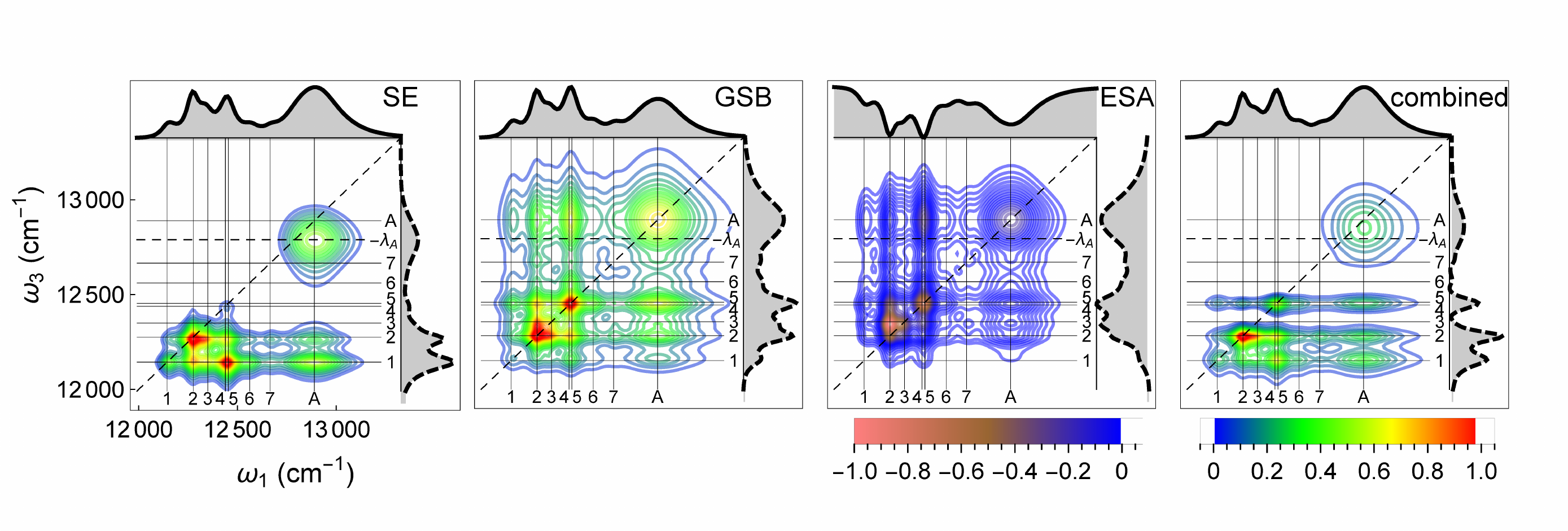}
\end{center}
\caption{2DES contributions at $\tau=100$~ps and $T=150$~K. 
Spectra obtained by integrating along the excitation and emission frequencies are shown on the figures top (solid lines) and right-hand sides (dashed lines).
\label{fig:2DESpathways}}
\end{figure}

The 2DES excitation spectrum superimposes three different excitation mechanisms, namely stimulated emission (SE), ground state bleaching (GSB) and excited state absorption (ESA), see Fig.~\ref{fig:2DESpathways}.
The interpretation of time-scales seen in 2DES is complicated by the different peak positions and dynamics underlying the different contributions. 
Only the SE part tracks directly the excitonic energy transfer towards the equilibrium state, while GSB contributes with a static background signal resembling the linear absorption spectra.
The ESA process is also unique to 2DES, where the pulse sequence leads to a further excitation of a second exciton.
This results in a negative signal, compared to the GSB and SE contributions.
Grey areas at each panel show the integrated 2D spectra over the excitation $\omega_1$ and emission frequency  $\omega_3$.
At large delay times, the top line of the combined signal (Fig.~\ref{fig:2DESpathways}) represents the absorption spectra and the lateral dashed line represents the transient absorption spectra. 
We observe in Fig.~\ref{fig:2DESpathways} that the SE contribution, which represents the dynamical energy transfer towards the thermal equilibrium state, shows a similar contribution of $(A, 2)$ with respect to $(A, 1)$. 
This is not expected at the lower cryogenic temperatures in the experiments \cite{Dostal2016}.
Note that inclusion of the RC \cite{Dostal2016} into our model would also mainly affect the relative weights of the cross-peaks $(A, i=1,2)$ in the SE spectra, which are determined by the thermal population weights.
Additionally we observe how the $(A, i=4,5)$ cross-peaks in the combined signal (Fig.~\ref{fig:2DESpathways}) are absent in the individual SE contribution. 
They are a feature of the GSB and ESA contributions in Fig.~\ref{fig:2DESpathways} and remain constant for long delay times, see Fig.~\ref{fig:all2DES}. 
The most prominent feature in systems with larger reorganization energies, such as the chlorosome antenna, is the shift of the SE $(A,A)$ peak from the diagonal of the 2DES to the bare exciton energy (Fig.~\ref{fig:2DESpathways}).
With the addition of the GB and ESA parts, which are not undergoing this shift, the combined 2DES (Fig.~\ref{fig:2DESpathways}) features a shift of about half the reorganization energy.
From the difference of the excitation and emission frequencies in the 2DES observed at the chlorosome peak \cite{Dostal2016}, we estimate the reorganizational shift of the chlorosome antenna to be $\lambda_A \sim 300-400$~cm$^{-1}$, which is in line with theoretical predictions \cite{Fujita2012}
and inline with our absorption spectra calculation in Fig.~\ref{fig:ctepschema}.

\begin{figure}[t]
\begin{center}
\includegraphics[width=0.6\textwidth]{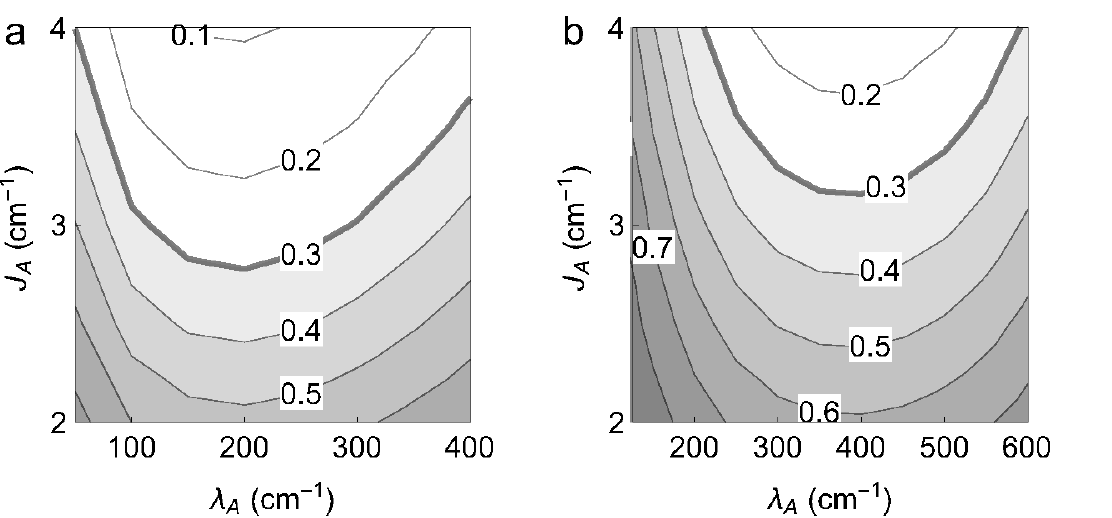}
\end{center}
\caption{
Chlorosome antenna population after $60$~ps at temperature $T=300$~K for different values of the coupling $J_{A}$ between the antenna and FMO BChl~1 and BChl~6 and reorganization energy of the antenna $\lambda_A$ for (a) the red-shifted antenna peak position ($S_A=12940$~cm$^{-1}$) and for (b) the unshifted antenna peak at
$S_A^0=13300$~cm$^{-1}$.
For fixed $J_A$, the fastest transfer time is reached for $S_A^0$ at $\lambda \sim 400$~cm$^{-1}$.
\label{fig:optilambda}}
\end{figure}

Next we establish the effective coupling $J_A$ between the FMO unit and the chlorosome antenna with its larger reorganization energy $\lambda_A$ .
The chosen coupling strength $J_A$ is determined from a comparison with the experimentally recorded  time scales of the exciton relaxation, namely $\sim 70$~ps scale for the antenna decay into the reaction center \cite{Dostal2016}. 
The transfer time, seen as population decrease of the antenna, becomes shortest for specific conjunctions of reorganization energies and coupling strengths.
We show the population transfer after $60$~ps for two different laser spectra, the redshifted one used in the 2DES calculations and the unshifted one, in Fig.~\ref{fig:optilambda}. 
Increasing the coupling $J_A$ leads to monotonous decrease of the chlorosome antenna population, while changing the reorganization energy for fixed coupling gives the optimal value of $\lambda_A$ for an efficient depletion \cite{Kreisbeck2011}.
For the broad band laser covering all chlorosome energies, an optimal value $\lambda^0_A\sim 300-400$~cm$^{-1}$ leads to the fastest inter-unit transfer in Fig.~\ref{fig:optilambda}b.
For the red-shifted laser and antenna peak, a similar transfer efficiency is achieved for $\lambda_A\sim 100-200$~cm$^{-1}$ in Fig.~\ref{fig:optilambda} a.
In both cases, the coupling $J_A=3$~cm$^{-1}$ matches the experimentally observed depletion of the chlorosome antenna population to $0.3$ after $60$~ps.  
Here, we have chosen $J_A=3$~cm$^{-1}$ and $\lambda_A=100$~cm$^{-1}$ for the 2DES simulation.
The decay time of the chlorosome antenna population $p(A)={\rm e}^{-t/t_{\rm decay}}$ in Fig.~\ref{fig:optilambda} is only weakly affected by temperature ($t_{\rm decay,300K}=52.5$~ps, $t_{\rm decay,100K}=57$~ps).

\begin{figure}[t]
\begin{center}
\includegraphics[width=0.4\textwidth]{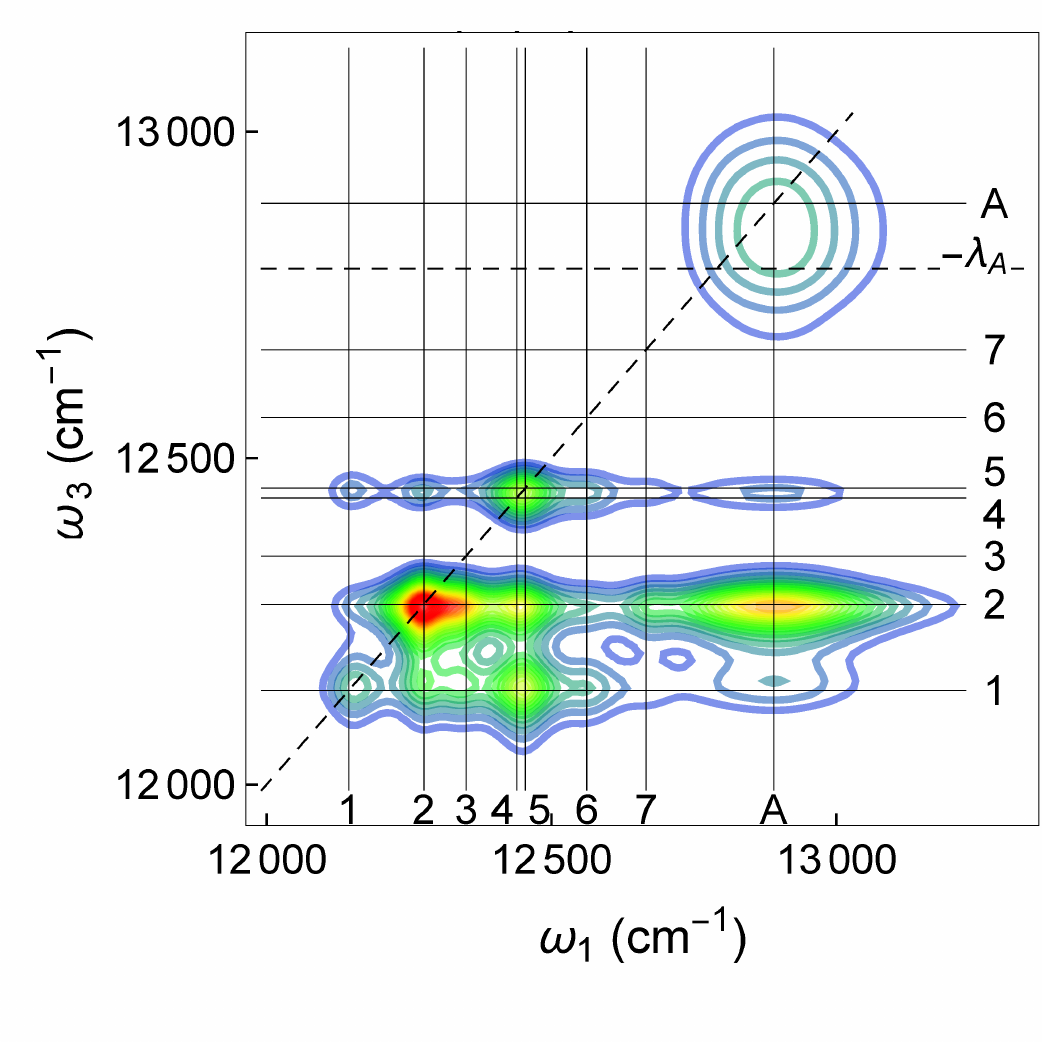}
\end{center}
\caption{2DES at $\tau=100$~ps for a perpendicular orientation of the antenna and the baseplate plane.
The cross-peak (A,1) is diminished compared to the parallel orientation used in Fig.~\ref{fig:all2DES}.
\label{fig:dipor}}
\end{figure}

In addition to the position of the spectral peaks and the excitonic couplings, 2DES is sensitive to the dipole orientation.
The FMO complex dipoles are oriented along the $N_B$, $N_D$ nitrogen atoms of the BChl PDB structure \cite{Tronrud2009,Muh2007}.
Chemical labelling and mass spectrometry data have established the orientation of the FMO protein on the baseplate membrane \cite{Wen2009} and the chlorosome \cite{Orf2013} as shown in Fig.~\ref{fig:ctepschema}. 
For the tubular antenna complex we orient the effective dipole parallel to the baseplate.
If instead a perpendicular antenna-baseplate orientation is chosen, the 2DES calculations differ and the cross peak between the chlorosome antenna and lowest exciton state $1$  is diminished, Fig.~\ref{fig:dipor}.
The lowest exciton state $1$, located mainly on BChl~3, is in particular diminished by the perpendicular antenna orientation with respect to the baseplate due to the BChl~3 dipole orientation parallel to the baseplate plane (see Fig.~\ref{fig:ctepschema}).
To account for different rotations of the antenna within the baseplate plane, we average over two orthogonal configurations (Tab.~\ref{tab:param}). 

\section*{Methods}

\begin{table}[b]
\begin{center}
\begin{tabular}{|l|r|}
\hline
reorganization energy $\lambda_A$ ($\lambda_A^0$) & $100\; (300)$~cm$^{-1}$  \\
coupling to BChl~1 and 6 ~~ $J_{A}$      & $3$~cm$^{-1}$ \\
peak absorption         $S_A$ ($S_A^0$)           &  $12940\; (13300)$~cm$^{-1}$ \\
dipole strength         $\mu_A$($\mu^0_A$)        &  $2.1\;(8.4)$~$\times\mu_\text{FMO}$ \\
antenna orientation (parallel)     $\hat{\mu}_{A,1}$, $\hat{\mu}_{A,2}$ & $(0.208, -0.788, 0.580)$, $(-0.790, 0.215, 0.575)$ \\
perpendicular         $\hat{\mu}_{A,\perp}$   & $(1,1,1)/\sqrt{3}$ \\\hline
\end{tabular}
\end{center}
\caption{Parameter of the coupled antenna chlorosome / FMO complex. $S_A^0$,
$\lambda_A^0$ and $\mu_A^0$ have been used for absorption spectra calculations in Fig.~\ref{fig:ctepschema}. 
The dipole orientations are given with respect to the protein data bank 3ENI FMO structure.
\label{tab:param}
}
\end{table}

\paragraph{Exciton parameters}
The energy transport in the photosynthetic system is modeled using the Frenkel exciton description \cite{May2004}, including the external electromagnetic field from the laser pulses $H_{\rm field}(t)$ in the impulsive limit \cite{Mukamel1995}.
Besides the seven single exciton states of the FMO system and the antenna chlorosome,
additionally 28 two exciton states are explicitly included in the Hamiltonian and in the dipole matrix \cite{Cho2005,Hein2012}.
The excitonic site energies and couplings are taken from Adolphs and Renger \cite{Adolphs2006a}, (Tab.~4, col.~2 and Tab.~1 col.~4 respectively).
The antenna is added as an additional pigment with energy in the chlorosome region $S_A=12940$~cm$^{-1}$ (see Fig.~2 in Ref.~\citenum{Dostal2016}).
This artificially red-shifted energy increases the FMO signal compared to the otherwise dominant absorbance of the chlorosome.
Only for the linear absorption calculation (measured with a broader frequency laser covering the complete chlorosome spectrum, see Fig.~1 in Ref.~\citenum{Dostal2016}) we take $S_A^0=13300$~cm$^{-1}$.
The chlorosome is linked to the FMO via the baseplate located on top of BChl~1,  BChl~6 and  BChl~8 (Fig.~\ref{fig:ctepschema}a).
The baseplate is set parallel to the plane of  the Mg center atoms of the BChl 8 pigments of the trimeric FMO complex.
BChl~6 and BChl~1 have been considered before as entry channel for energy flow from the chlorosome \cite{Wen2009}.
The more recent discovery of an eight chlorophyll in proximity of the baseplate and BChl~1 also suggest that this pigment links the chlorosome to the FMO \cite{Kell2016,Thyrhaug2016}.
Both, the baseplate and the eight pigment have not been observed in 2DES \cite{Dostal2016}.
Therefore we do not include them in the model, but directly couple the chlorosome to the FMO BChl~1 and BChl~6. 
For the vibrational couplings of the BChls and the chlorosome, we consider eight independent sets of harmonic oscillators coupled linearly to each pigment.
The spectral density of the vibrations of the $m$th pigment is given by Drude-Lorentz shape $J_m(\omega)=2\lambda_m\frac{\omega\gamma}{\omega^2+\gamma^2}$ with a correlation time $\gamma^{-1}=50$~fs.
We set for the seven FMO BChls $\lambda_{\rm FMO}=35$~cm$^{-1}$.

\paragraph{2DES calculation}
2DES signals are obtained from a sequence of  three ultra-short laser pulses and from recording the resulting signal at a fixed time delay $\tau$ between the second and third pulse.
The resulting signal is proportional to the third order optical response function $S^{(3)}(\omega_3,\tau,\omega_1)$, where 
the excitation frequency is denoted by $\omega_1$ and the emission frequency by $\omega_3$.
The frequency resolved 2DES are computed from a time propagation of the density matrix using the numerically exact hierarchical equation of motion (HEOM) method on massively parallel graphic processing units \cite{Kreisbeck2011,Hein2012,Kreisbeck2014}.
In contrast to perturbative treatments using Redfield or F\"orster theory, HEOM accounts for the different reorganization process of the FMO complex compared to the chlorosome antenna.
With increasing reorganization energy, the computational and memory requirements for HEOM increase, in particular for computing the two exciton states. 
By using the high temperature approximation at $150$~K and truncating  $\lambda_A=100$~cm$^{-1}$  we ensure converged results of all simulations and capture the larger reorganizational shift of the antenna.
The effect of disorder is not directly considered here, but has been discussed before for various disorder parameters of the FMO complex \cite{Hein2012} and does not affect the presented analysis.
2DES are often measured for two different laser setups, stemming from the rephasing and non-rephasing contributions.
All spectra are rotationally averaged over 20 different orientations of the joint complex with respect to the laser polarization \cite{Hein2012}.
In addition we average over two orthogonal orientations of the chlorosome antenna in the baseplate plane.

\section*{Conclusion}

By modeling 2DES of the photosynthetic unit of {\it C.~tepidum}, we have studied energy transfer processes between the chlorosome and the FMO complex.
We show how the energy funnels from higher to lower energies not only within the FMO complex but also within subunits of the photosynthetic apparatus.
The structural variety of the subunits in the photosynthetic apparatus of  {\it C.~tepidum} leads to different reorganisational strengths.
To account for these differences, we use the HEOM formalism which is required to properly model the varying environmental couplings and time-scales \cite{Kreisbeck2015,Olsina2014}.
We show excellent agreement with the main features of the experimental 2DES \cite{Dostal2016} and 
demonstrate that energy transfer processes occur on a slow scale of tens of picoseconds between the subsystems.
The transfer time is still fast compared to the nanoseconds lifetime of excitons in isolated chlorosomes \cite{Blankenship2014} and allows the photosynthetic apparatus to function with high efficiency.

From comparison with the experimental data, we estimate the effective reorganization energy of the chlorosome antenna, the inter unit coupling, and the relative orientation between subsystems.
We find that the energy is transferred from the antenna chlorosome to the FMO complex via the strong environmental coupling, but with only weak dipole-dipole interactions between the subunits.
The strong reorganisational energies in the chlorosome result in ranges of couplings which optimise the transfer time between the subunits.

\section*{Acknowledgements}
The work was supported by the North-German Supercomputing Alliance (HLRN) and by the German Research Foundation (DFG) grant KR~2889. 
M.R.\ has received funding from the European Union's Horizon 2020 research and innovation programme under the Marie Sklodowska-Curie grant agreement No~707636.

\section*{Author contributions statement}

T.K. and M.R. contributed equally to the theoretical analysis and writing of the manuscript.

\section*{Additional information}

The authors declare no competing financial interests.

\end{document}